\documentclass[conference,a4paper]{IEEEtran}
\IEEEoverridecommandlockouts
\usepackage{cite}
\usepackage[cmex10]{amsmath}
\usepackage{amssymb,amsfonts}
\usepackage{algorithmic}
\usepackage{graphicx}
\usepackage{textcomp}
\usepackage{xcolor}
\def\BibTeX{{\rm B\kern-.05em{\sc i\kern-.025em b}\kern-.08em
    T\kern-.1667em\lower.7ex\hbox{E}\kern-.125emX}}

\usepackage[utf8]{inputenc}
\usepackage[a4paper, total={17cm,25cm}]{geometry}
\usepackage{mleftright}       %
\mleftright                   %

\usepackage{amsthm}
\usepackage{listings}
\usepackage{pgfplots}
\usetikzlibrary{positioning}

\usetikzlibrary{arrows.meta}
\usetikzlibrary{backgrounds}
\usepgfplotslibrary{patchplots}
\usepgfplotslibrary{fillbetween}
\pgfplotsset{%
layers/standard/.define layer set={%
    background,axis background,axis grid,axis ticks,axis lines,axis tick labels,pre main,main,axis descriptions,axis foreground%
}{grid style= {/pgfplots/on layer=axis grid},%
    tick style= {/pgfplots/on layer=axis ticks},%
    axis line style= {/pgfplots/on layer=axis lines},%
    label style= {/pgfplots/on layer=axis descriptions},%
    legend style= {/pgfplots/on layer=axis descriptions},%
    title style= {/pgfplots/on layer=axis descriptions},%
    colorbar style= {/pgfplots/on layer=axis descriptions},%
    ticklabel style= {/pgfplots/on layer=axis tick labels},%
    axis background@ style={/pgfplots/on layer=axis background},%
    3d box foreground style={/pgfplots/on layer=axis foreground},%
    },
}

\usepackage{xspace}
\usepackage[ruled,vlined]{algorithm2e}
\usepackage{tabularx}
\usepackage{booktabs}
\usepackage{arydshln}
\usepackage[normalem]{ulem}
\usepackage{placeins}
\usepackage{balance}

\definecolor{DarkGreen}{rgb}{0.1,0.5,0.1}
\definecolor{DarkRed}{rgb}{0.5,0.1,0.1}
\definecolor{DarkBlue}{rgb}{0.1,0.1,0.5}

\usepackage[pdftex]{hyperref}
\hypersetup{
    unicode=false,          %
    pdftoolbar=true,        %
    pdfmenubar=true,        %
    pdffitwindow=false,      %
    pdfnewwindow=true,      %
    colorlinks=true,       %
    linkcolor=DarkBlue,          %
    citecolor=DarkGreen,        %
    filecolor=DarkGreen,      %
    urlcolor=DarkBlue,          %
    pdftitle={Trading Communication for Computation in Byzantine-Resilient Gradient Coding},
    pdfauthor={},
}
\usepackage[capitalize]{cleveref}
\usepackage{siunitx}
\usepackage{intcalc}

\usepackage[inline]{enumitem}

\newtheorem{theorem}{Theorem}
\newtheorem{corollary}{Corollary}

\newtheorem{definition}{Definition}

\newtheorem{example}{Example}
\newtheorem{remark}{Remark}

\DeclareMathOperator*{\argmin}{arg\,min}
\newcommand{\lis}[1]{\ensuremath{\mathfrak{#1}}} 
\newcommand{\setstyle}[1]{\ensuremath{\mathcal{#1}}} 
\newcommand{\compbound}{\ensuremath{\left\lfloor \frac{\nmalicious}{\nhonest}\right\rfloor}\xspace}
\newcommand{\nworker}{\ensuremath{n}\xspace}
\newcommand{\ngroup}{\ensuremath{m}\xspace}
\newcommand{\ngrad}{\ensuremath{p}\xspace} 
\newcommand{\graddim}{\ensuremath{d}\xspace}
\newcommand{\tgrad}{\ensuremath{\mathbf{g}}\xspace} 
\newcommand{\gestim}{\ensuremath{\mathbf{\widehat{g}}}\xspace} 
\newcommand{\pgrad}[1][\gind]{\ensuremath{\mathbf{g}_{#1}}\xspace} 
\newcommand{\wpgrad}[2]{\ensuremath{{\mathbf{\widetilde{g}}^{(#2)}_{#1}}}\xspace} 
\newcommand{\exwpgrad}[2]{\ensuremath{\widetilde{g}^{(#2)}_{#1}}\xspace} 
\newcommand{\gind}{\ensuremath{i}\xspace}
\newcommand{\wind}{\ensuremath{j}\xspace}
\newcommand{\compind}{\ensuremath{\zeta}\xspace} 
\newcommand{\sample}[1][\gind]{\ensuremath{\mathbf{x}_{#1}}\xspace}
\newcommand{\worker}[1][\wind]{\ensuremath{W_{#1}}\xspace}
\newcommand{\galpha}{\ensuremath{\setstyle{A}}\xspace} 
\newcommand{\alphasize}{\ensuremath{{\card{\galpha}}}\xspace} 
\newcommand{\nmalicious}{\ensuremath{s}\xspace}
\newcommand{\encfun}[1][]{\ensuremath{\operatorname{enc}_{#1}}\xspace}
\newcommand{\encind}{\ensuremath{e}\xspace}
\newcommand{\encfunset}{\ensuremath{\lis{E}}\xspace}
\newcommand{\decfun}[1][]{\ensuremath{\operatorname{dec}_{#1}}\xspace}
\newcommand{\nencfun}{\ensuremath{v}\xspace} 
\newcommand{\replfact}[1][]{\ensuremath{\rho_{\textrm{#1}}}\xspace}
\newcommand{\commoh}[1][]{\ensuremath{\kappa_{\textrm{#1}}}\xspace}
\newcommand{\proto}{\ensuremath{\Pi}\xspace}
\newcommand{\nround}[1][]{\ensuremath{r_{#1}}\xspace} 
\newcommand{\localcomp}[1][]{\ensuremath{c_{#1}}\xspace} 
\newcommand{\elimset}{\ensuremath{{\setstyle{S}}}\xspace}
\newcommand{\params}{\ensuremath{\boldsymbol{\theta}}\xspace}
\newcommand{\learningrate}{\ensuremath{\eta^{(\gditer)}}\xspace}
\newcommand{\loss}{\ensuremath{\ell}\xspace}
\newcommand{\gditer}{\ensuremath{\tau}\xspace} 
\newcommand{\range}[1]{\ensuremath{[#1]}\xspace}
\newcommand{\roundind}{\ensuremath{t}\xspace}
\newcommand{\respdim }[2]{\ensuremath{d_{{#2},{#1}}}\xspace}
\newcommand{\iresp}{\ensuremath{\mathbf{z}}\xspace}
\newcommand{\noisyiresp}{\ensuremath{\mathbf{\widetilde{z}}}\xspace}
\newcommand{\irespset}{\ensuremath{\lis{Z}}\xspace}
\newcommand{\recresset}{\ensuremath{\widetilde{\irespset}}\xspace}
\newcommand{\workerset}{\ensuremath{\setstyle{W}}\xspace}
\newcommand{\locgradset}{\ensuremath{\lis{G}}\xspace}
\newcommand{\lgindset}{\ensuremath{\lis{I}}\xspace}
\newcommand{\disagreegradset}{\ensuremath{\setstyle{\widetilde{I}}}\xspace}
\newcommand{\allocmat}[1][]{\ensuremath{\mathbf{A}_{#1}}}
\newcommand{\wtmdata}{\ensuremath{\setstyle{D}}\xspace}
\newcommand{\I}{\ensuremath{\mathrm{I}}\xspace}
\newcommand{\En}{\ensuremath{\mathrm{H}}\xspace}
\newcommand{\given}{\;\!\vert\;\!}
\newcommand{\card}[1]{{\left\vert #1 \right\vert}}
\newcommand{\defeq}{\overset{\text{\tiny def}}{=}}
\newcommand{\master}{main node\xspace}
\newcommand{\BGC}{$\nmalicious$-BGC\xspace}
\newcommand{\R}{\mathbb{R}}
\newcommand{\N}{\mathbb{N}}
\newcommand{\playA}{Alice\xspace}
\newcommand{\playB}{Bob\xspace}
\newcommand{\playC}{Carol\xspace}
\newcommand{\playboss}{Dan\xspace}
\newcommand{\playbossletter}{D\xspace}
\newcommand{\playbosspronoun}{he\xspace}
\newcommand{\playbossPronoun}{He\xspace}
\newcommand{\nhonest}{\ensuremath{u}}
\newcommand{\indone}{{\wind_1}}
\newcommand{\indtwo}{{\wind_2}}
\newcommand{\indcheck}{{\gind_\mathrm{check}}}
\newcommand{\comp}[2][\compind]{{\left[ {#2} \right]_{#1}}}

\renewcommand{\baselinestretch}{0.946}

\newif\ifarxiv 
\newif\ifisit

\arxivtrue
\isitfalse

\newcommand{\ARXIVonly}[1]{%
  \ifarxiv%
    #1%
  \fi%
}
\newcommand{\ISITonly}[1]{%
  \ifisit%
    #1%
  \fi%
}

\newcommand{\Revision}[1]{#1}

\newcommand{\RevisionRemove}[1]{}

\begin{document}

\title{Trading Communication for Computation in Byzantine-Resilient Gradient Coding}

\author{
	\IEEEauthorblockN{Christoph Hofmeister, Luis Maßny, Eitan Yaakobi, and Rawad Bitar}
 \vspace{-1cm}
    \thanks{CH, LM and RB are with the School of Computation, Information and Technology at the Technical University of Munich, Germany. Emails: \{christoph.hofmeister, luis.massny, rawad.bitar\}@tum.de}
    \thanks{EY is with the CS department of Technion---Israel Institute of Technology, Israel. Email: yaakobi@cs.technion.ac.il}
    \thanks{This project is funded by the Technical University of Munich - Institute for Advanced Studies, funded by the German Excellence Initiative and European Union Seventh Framework Programme under Grant Agreement No. 291763, by the Bavarian Ministry of Economic Affairs, Regional Development and Energy within the scope of the 6G Future Lab Bavaria, and by DFG (German Research Foundation) projects under Grant Agreement No. WA 3907/7-1 and No. BI 2492/1-1.}
}

\maketitle

\begin{abstract}
  We consider gradient coding in the presence of an adversary\RevisionRemove{,} controlling so-called malicious workers trying to corrupt the computations. Previous works propose the use of MDS codes to treat the inputs of the malicious workers as errors and correct them using the error-correction properties of the code. This comes at the expense of increasing the replication, i.e., the number of workers each partial gradient is computed by. In this work, we reduce replication by proposing a method that detects the erroneous inputs from the malicious workers, hence transforming them into erasures. For $\nmalicious$ malicious workers, our solution can reduce the replication to $\nmalicious+1$ instead of $2\nmalicious+1$ for \emph{each partial gradient} at the expense of only $\nmalicious$ additional computations at the \master and additional rounds of light communication between the \master and the workers. We give fundamental limits of the general framework for fractional repetition data allocation.
  Our scheme is optimal in terms of replication and local computation but incurs a communication cost that is asymptotically, in the size of the dataset, a multiplicative factor away from the derived bound.
\end{abstract}

\section{Introduction}

Consider the setting of a \master possessing large amounts of data on which a machine learning model shall be trained using gradient descent. To speed up the learning process, the \master distributes the computations to several worker nodes~\cite{lian2017can,abadi2016tensorflow}.
One of the main vulnerabilities of distributed gradient descent is the presence of Byzantine errors corrupting some workers' computation results~\cite{lamportByzantineGeneralsProblem}. Even a single corrupted computation result can drastically deteriorate the performance of the algorithm~\cite{damaskinos2019aggregathor}.

The problem of tolerating Byzantine errors in distributed computing has been considered in different settings. For example, for linear computations\RevisionRemove{, i.e., matrix-matrix or matrix-vector multiplications,}~\cite{hofmeisterSecurePrivateAdaptive2022,tangAdaptiveVerifiableCoded2022} use Freivalds' algorithm to detect Byzantine errors with high probability and exclude them in further processing. For polynomial computations, \cite{yuLagrangeCodedComputing2019} uses properties of error-correcting codes to correct \Revision{errors}\RevisionRemove{the erroneous results}. Other approaches towards the mitigation of erroneous results include \RevisionRemove{the use of} group testing and Reed-Solomon codes~\cite{solankiNonColludingAttacksIdentification2019}, and homomorphic hash functions~\cite{keshtkarjahromiSecureCodedCooperative2019}. For a more comprehensive review of the existing literature, we refer the interested reader to~\cite{hofmeisterSecurePrivateAdaptive2022}. %

For distributed gradient descent each worker computes a gradient of a so-called loss function for local training data. The \master aggregates \RevisionRemove{the locally computed}\Revision{these partial} gradients into a total gradient. The problem of Byzantine errors in this context has been first tackled using robust aggregation functions, which select only a subset of the workers' results. The selection is based on minimizing the distance to other results~\cite{guerraoui2018hidden,chenDistributedStatisticalMachine2017,blanchardMachineLearningAdversaries,rajputDETOXRedundancybasedFramework,damaskinos2019aggregathor,el2020fast}, by assigning a sanity score to each result~\cite{xieZenoDistributedStochastic}, or by general consistency checks~\cite{konstantinidisRobustDetection}. Using such aggregation functions, however, the resulting gradient estimate may be only an inexact approximation of the desired total gradient in the error-free case. This increases the runtime for the gradient descent algorithm, see e.g.~\cite{bitarStochasticGradientCoding2020} and references therein, and might perform poorly in some particular settings, e.g., when the distribution of the training data is not identical among the workers~\cite{chenRevisitingDistributedSynchronous2017, tandonGradientCoding2017}. 
Moreover, advanced gradient descent techniques, such as the momentum method~\cite{sutskeverImportanceInitializationMomentum}, are in general not compatible with approximate schemes~\cite{tandonGradientCoding2017}.

Due to the latter, the problem of tolerating Byzantine errors in distributed gradient descent with exact recovery has been approached from a coding-theoretic perspective. For linear and polynomial computations, coding over the input data at the \master has been proposed in~\cite{yuLagrangeCodedComputing2019} and~\cite{dataDataEncodingByzantineResilient2021}. Since the computations in gradient descent are highly non-linear in general, these approaches might not be applicable. Instead,\RevisionRemove{ the work of}~\cite{chenDRACOByzantineresilientDistributed2018} introduces DRACO, a framework that performs coding over the computation results at the workers. The authors build on the idea of gradient coding~\cite{tandonGradientCoding2017}, which \RevisionRemove{had}\Revision{was} originally\RevisionRemove{ been} designed to mitigate the effect of stragglers, i.e., slow or unresponsive workers. By replicating each gradient computation to $\nmalicious+1$ different workers, gradient coding can tolerate $\nmalicious$ stragglers by treating them as erasures. Applying the same ideas, DRACO can tolerate $\nmalicious$ malicious workers instead, treating the computations of malicious workers as errors. This comes at the cost of increasing the replication of each gradient computation to $2\nmalicious+1$, hence, causing a large computation overhead. Both DRACO and gradient coding are shown to achieve an optimal replication for the respective problem settings.

In this work, we also consider the problem of exact gradient coding in the presence of an adversary controlling $\nmalicious$ malicious workers, that introduce Byzantine errors in their computation results.
In contrast to~\cite{chenDRACOByzantineresilientDistributed2018}, we define a more general framework in which the \master can run a small number of computations itself to aid in decoding. We propose a scheme that requires a replication of only $\nmalicious+1$ at the expense of running $\nmalicious$ local gradient computations\footnote{A replication of $\nmalicious+1$ means that \Revision{\emph{each}} gradient computation is run $\nmalicious+1$ times; the cost of $\nmalicious$ local computations \Revision{\emph{in total}} is very small in comparison.}. The idea of running local computations at the \master is also used in~\cite{caoDistributedGradientDescent2019,prakash2020secure}, where the \master computes an estimate of the true gradient from only few data samples and discards worker results that have a large distance to this estimate. 
In contrast to our work, those solutions do not guarantee exact recovery of the total gradient at the \master. We design additional light communication between the workers and the \master to help identify which gradients should be computed locally. We generalize the framework to the case where less than $\nmalicious$ local computations are allowed at the \master and explore the tradeoff between replication and local computation.

\vspace{-1em}
\section{Problem Setting}
We first set the notation. 
Matrices and vectors are denoted by upper-case and lower-case bold letters, respectively. $\mathbf{A}_{i,j}$ refers to the element in row $i$ and column $j$ of the matrix $\mathbf{A}$. Scalars are denoted by lower-case letters, sets by calligraphic letters, and lists by fractal letters, respectively, e.g., $a$, $\mathcal{A}$ and $\lis{A}$. For an integer $a \geq 1$, we define $\range{a} \defeq \left\{ 1,2,\dots,a \right\}$.
Let $\lis{A}_i, \, i=1,\dots,t$, be a collection of lists, we define $\lis{A}^{(t)}$ to be their concatenation.
We use $\boldsymbol{1}_{m \times n}$ and $\boldsymbol{0}_{m \times n}$ to denote the all-one and all-zero matrices of dimension $m \times n$. 
\ARXIVonly{For a list of symbols, cf. \cref{app:notations}.}

We consider a \emph{synchronous distributed gradient descent} setting, in which the goal is to fit the parameters $\params \in \R^\graddim$ of a model to a dataset consisting of $\ngrad$ samples $\sample \in \R^\graddim, \gind \in
\range{\ngrad}$. This is done by finding (local) optima for the problem
$\displaystyle \argmin_{\params \in \R^\graddim} \sum_{\gind \in \range{\ngrad}} \loss(\params, \sample)$ 
for a per-sample loss function $\loss(\params, \sample)$.
The gradient descent algorithm starts with a random initialization for the 
parameter vector, defined as $\params^{(0)}$, and then iteratively applies the update rule $
    \params^{(\gditer+1)} = \params^{(\gditer)} - \frac{\learningrate}{\ngrad} \sum_{\gind \in
    \range{\ngrad}} \nabla \loss(\params^{(\gditer)}, \sample),
$ where $\gditer$ is the iteration index and $\learningrate \in \R$ is referred to as the learning rate. For notational convenience, we define the evaluation of the gradient of the loss function at individual samples as $
    \pgrad^{(\gditer)} \defeq \nabla \loss(\params^{(\gditer)}, \sample)
$ and call them \emph{partial gradients}.
\Revision{%
Since in practice data 
is quantized to a finite set of values, we 
take the partial gradients to be
vectors over a finite alphabet $\galpha$, i.e., $\pgrad^{(\gditer)} \in \galpha^\graddim$.}

Consider a system comprising a \master and $\nworker$ worker nodes, $\nmalicious$ of which might be malicious.
The malicious workers can send arbitrarily corrupted information to the \master. At the start of the procedure, the \master distributes the samples to the workers with some redundancy.
Then, each iteration $\gditer$ starts with the \master broadcasting the current parameter vector $\params^{(\gditer)}$. The workers then compute the partial gradients $\pgrad^{(\gditer)}$ corresponding to the samples they store. At the end of the iteration, the \master must obtain the \emph{full gradient} $\tgrad^{(\gditer)} \defeq \sum_{\gind \in \range{\ngrad}} \pgrad^{(\gditer)}$ irrespective of the actions of the $\nmalicious$ malicious workers.

In this work, we are concerned with the problem of reliably reconstructing $\tgrad^{(\gditer)}$ \emph{exactly} at the \master in each iteration $\gditer$.
In the sequel, we only consider a single iteration of gradient descent and omit the superscript $\gditer$.

\section{Byzantine-Resilient Gradient Coding}
For a better grasp of our ideas, we start with a toy example that captures the concepts introduced and studied.

\subsection{How to Catch Liars Efficiently?}
Consider a game among three friends \playA (A), \playB (B) and \playboss (D). A and B have $\ngrad$ private numbers $g_1,\dots, g_{\ngrad}$ whose sum should be communicated correctly to \RevisionRemove{\playboss}\Revision{D}. The problem is that one player is trying to cheat \RevisionRemove{on} D.

At the first stage, each player sends the sum of all numbers to \playbossletter. Then, the game is played in rounds. At each round, \playbossletter can first ask A and B questions about $g_1,\dots,g_{\ngrad}$. Only one player is guaranteed to reply truthfully. Then, \playbossletter can query an oracle to uncover the true value of some of the $g_i$'s. The game is repeated until \playbossletter correctly obtains the desired sum.
 
\begin{example}
  Assume w.l.o.g that A is cheating \RevisionRemove{on} \playbossletter, $\ngrad=4$, and that $g_i = i$. In a first stage, A and B send the sum of their numbers to \playbossletter. Assume that A sends the value $2$ and B sends the correct value $10$. At the first round, \playbossletter asks A and B to send the sum $g_1+g_2$. To create confusion, A acts truthfully and also sends the value $3$. Now, \playbossletter knows that $g_3+g_4$ is either $-1$ or $8$. So, \playbossletter decides not to query the oracle yet. He instead moves to the second round and asks A and B to send the value of $g_3$. Again, both players send the value $3$. Hence, \playbossletter now knows that $g_4$ has to be $-4$ if A is acting honestly or $5$ if B is acting honestly. \RevisionRemove{He then decides to query the oracle, obtain $g_4$, catch the liar A, and obtain the true sum.}
  \Revision{At this point, \playbossletter decides to query the oracle for $g_4$, thus catching the liar A and obtaining the true sum $10$.}

\label{ex:toy_problem}
\end{example}

This game is the crux of our framework. The private numbers are the \Revision{partial} gradients computed at the workers, querying the oracle represents local computations at the \master, and asking questions is the \RevisionRemove{introduced} light communication between the \master and the workers. The figures of merit of this game are: \begin{enumerate*}[label={\emph{\roman*)}}]
        \item the minimum number of queries to the oracle that \playbossletter needs; and 
        \item given that \playbossletter can only obtain the minimum number of oracle queries and can play several rounds, how many questions does \playbossletter need to ask the other players to recover the desired sum correctly.
\end{enumerate*}
For a more elaborate example see~\ARXIVonly{\cref{app:biggerexample}}\ISITonly{\cite[Appendix A]{hofmeisterBGC}}.

\vspace{-0.1cm}
\subsection{The Framework}

Carrying over this idea to the gradient coding framework, we next define gradient coding schemes resilient against an  adversary controlling $\nmalicious$ malicious workers.

\begin{definition}[Byzantine-resilient gradient coding scheme]
    A Byzantine-resilient gradient coding scheme tolerating $s$ malicious workers, referred to as \BGC, is a tuple $\left( \allocmat, \encfunset, \decfun, \proto \right)$ where
\begin{itemize}
    \item $\allocmat \in \{0, 1\}^{\ngrad \times \nworker}$ is a \textbf{data assignment matrix} in which $\allocmat[{\gind, \wind}]$ is equal to $1$ if the $\gind$-th data sample is given to the $\wind$-th worker and $0$ otherwise,
    \item $\encfunset \defeq \left(\encfun[\wind,\encind] \mid \wind\in \range{\nworker}, \encind \in \range{\nencfun}\right)$ is the list of $\nworker\nencfun$ \textbf{encoding functions} used by the workers such that $\encfun[\wind,1]$ corresponds to a gradient code dictated by $\allocmat$ and $\encfun[\wind,\encind]$ depends only on the gradients assigned to $\worker$, %
    \item $\proto = (\proto_1,\proto_2)$ is a \textbf{multi-round protocol} in which $\proto_1$ selects the indices of the encoding functions to be used by the workers and $\proto_2$ selects gradients to be locally computed at the \master,
    \item and $\decfun$ is a \textbf{decoding function} used by the \master after running the protocol $\proto$ to always output the correct full gradient if the number of malicious workers is at most $\nmalicious$.
\end{itemize}
\end{definition}

Each worker initially ($\roundind=0$)
sends a vector $\iresp_{0,\wind} \defeq \encfun[\wind,1]\left( \pgrad[1],\dots,\pgrad[\ngrad] \right) \in \galpha^{\graddim}$ that is a codeword symbol of a gradient code~\cite{tandonGradientCoding2017}. The protocol $\proto$ then runs for $\nround \in \N$ rounds.
In each round $\roundind \in \range{\nround}$, the \master uses $\proto_1$ to select an encoding function $\encfun[\wind,\encind_{\roundind,\wind}]$ for each worker $\worker$ and communicates its index $\encind_{\roundind,\wind}$ to the respective worker. 
Each worker $\worker$ then computes a response $\iresp_{\roundind,\wind} = \encfun[\wind,\encind_{\wind,\roundind}]\left( \pgrad[1],\dots,\pgrad[\ngrad] \right) \in \galpha^{\respdim{\wind}{\roundind}}$ for some $\respdim{\wind}{\roundind} \in \N_0$ and sends a vector $\noisyiresp_{\roundind,\wind} \in \galpha^{\respdim{\wind}{\roundind}}$ to the \master.
For honest workers $\noisyiresp_{\roundind,\wind} = \iresp_{\roundind,\wind}$, while for malicious workers, $\noisyiresp_{\roundind,\wind}$ may be chosen arbitrarily.

In every round, the \master uses $\proto_2$ to choose a set of partial gradients to compute locally. We denote the list of indices of the locally computed partial gradients in round $\roundind$ by $\lgindset_\roundind$ and the corresponding list of partial gradient values by $\locgradset_\roundind \defeq (\pgrad[\gind] \mid \gind \in \lgindset_\roundind )$.
Analogously, we define $\irespset_\roundind \defeq (\iresp_{\roundind, \wind} \mid \forall \wind \in [\nworker])$ and 
$\recresset_\roundind \defeq (\noisyiresp_{\roundind, \wind} \mid \forall \wind \in [\nworker])$.

The protocol $\proto_1$ selects the indices of the encoding functions to be used in the current round $t$ based on the received results and locally computed gradients from previous rounds.
After receiving the results in the current round, the \master uses $\proto_2$ to select the gradients to compute locally in this round, i.e.,
\vspace{-0.2em}
\begin{align}
    \encind_{1,\roundind}, \dots, \encind_{\nworker, \roundind} &= \proto_1\left(\roundind, \recresset^{(\roundind-1)}, \lgindset^{(\roundind-1)}, \locgradset^{(\roundind-1)} \right), \label{eq:encind}\\
    \lgindset_\roundind &= \proto_2\left(\roundind, \recresset^{(\roundind)}, \lgindset^{(\roundind-1)}, \locgradset^{(\roundind-1)} \right) \label{eq:indset}.
\end{align}
After round $\nround$, the main node computes an estimate $\gestim$ of $\tgrad$ using the decoding function 
    $\gestim = \decfun\left(\recresset^{(\nround)}, \lgindset^{(\nround)}, \locgradset^{(\nround)} \right).$
The total number of partial gradients computed at the \master is defined as $\localcomp \defeq \card{\lgindset^{(\nround)}}$.

A valid \BGC scheme must output $\gestim = \tgrad$ if the number of malicious workers is at most $\nmalicious$.

\subsection{Figures of Merit}
An \BGC scheme is evaluated by the maximum number of rounds $\nround$ and the maximum number of local computations $\localcomp$ required by $\proto$, and its replication factor and communication overhead, which we define next.

\begin{definition}[Replication factor and communication overhead]
The \textbf{replication factor} of an \BGC scheme is the average number of workers to which each sample is assigned, i.e.,
\vspace{-1em}
\begin{equation*}
    \replfact \defeq \frac{\sum_{\gind \in \range{\ngrad}, \wind \in \range{\nworker}} \allocmat[{\gind, \wind}]}{\ngrad}.
\end{equation*}

The \textbf{communication overhead} is the maximum number of symbols from $\galpha$ transmitted from the workers to the \master during $\proto$, i.e.,
\begin{equation*}
    \commoh \defeq {\sum_{\roundind \in \range{\nround}, \wind \in \range{\nworker}} \respdim{\wind}{\roundind}}.
\end{equation*}
\end{definition}

We say that a tuple $(\allocmat, \encfunset, \decfun, \proto)$ is a $(\nround, \localcomp, \replfact, \commoh)$-\BGC scheme if in the presence of at most $\nmalicious$ malicious workers, the scheme always outputs $\gestim = \tgrad$ by requiring at most $\nround$ communication rounds, at most $\localcomp$ local computations, and has replication factor $\replfact$ and communication overhead less than or equal to $\commoh$.

We study settings in which the number of workers $\nworker$ divides $\nmalicious+1$, i.e., $\nworker = \ngroup (\nmalicious + 1)$ for some integer $\ngroup$ and only consider balanced data assignments, i.e., every worker computes the same number of gradients.
We focus on the particular case of a fractional repetition data assignment~\cite{tandonGradientCoding2017}. That is, the \master partitions the workers into $\ngroup$ groups of size $\frac{\nworker}{\ngroup}$ each and assigns the same data samples to all workers within a group.
The data assignment matrix is constructed as 
\vspace{-0.8em}
\begin{equation}
\label{eq:fractional_repetition}
\allocmat = \begin{bmatrix}
        \boldsymbol{1}_{\frac{\ngrad}{\ngroup} \times \frac{\nworker}{\ngroup}} & \boldsymbol{0}_{\frac{\ngrad}{\ngroup} \times \frac{\nworker}{\ngroup}} & \dots & \boldsymbol{0}_{\frac{\ngrad}{\ngroup} \times \frac{\nworker}{\ngroup}} \\
        \boldsymbol{0}_{\frac{\ngrad}{\ngroup} \times \frac{\nworker}{\ngroup}} & \boldsymbol{1}_{\frac{\ngrad}{\ngroup} \times \frac{\nworker}{\ngroup}} &   \dots & \boldsymbol{0}_{\frac{\ngrad}{\ngroup} \times \frac{\nworker}{\ngroup}} \\
        \vdots & \vdots & \ddots & \vdots \\
        \boldsymbol{0}_{\frac{\ngrad}{\ngroup} \times \frac{\nworker}{\ngroup}} & \boldsymbol{0}_{\frac{\ngrad}{\ngroup} \times \frac{\nworker}{\ngroup}} & \dots & \boldsymbol{1}_{\frac{\ngrad}{\ngroup} \times \frac{\nworker}{\ngroup}}
\end{bmatrix}.
\end{equation}
  \vspace{-1.1em}

Since we focus on this particular data assignment, the initial worker responses are given by the sum of all computed gradients
\smash{$\iresp_{0,\wind} =
\sum_{
\substack{\gind \in \range{\ngrad} \\ \allocmat[\gind,\wind]=1}
} \pgrad[\gind]$}, which form a valid gradient code.

\section{Bounds and Code Constructions of \BGC Schemes}
For the non-trivial case of $\localcomp<\ngrad$, i.e., the \master does not compute all the partial gradients locally, the replication factor of any \BGC scheme is bounded from below by %
$\replfact \geq {\nmalicious+1}$.
In addition, we note that if $\localcomp = 0$, then the replication factor of any \BGC scheme is bounded from below as $\replfact \geq {2\nmalicious+1}$ and can be achieved through DRACO \cite{chenDRACOByzantineresilientDistributed2018}. Conversely, if $\replfact \geq 2\nmalicious+1$, then $\localcomp=\commoh=0$ was shown to be achievable. Thus, we \Revision{focus on}\RevisionRemove{are most interested in} the case $\nmalicious+1 \leq \replfact \leq 2\nmalicious+1$, and we investigate the fundamental tradeoff between $\localcomp$, $\replfact$ and $\commoh$ for any $\nround$. 

In particular, we show that for $\localcomp = \nmalicious$ the replication factor $\replfact = {\nmalicious+1}$ is achievable. For $0<\localcomp<\nmalicious$, we show that for any $1 \leq \nhonest \leq \nmalicious+1$, if $\replfact \leq {\nmalicious + \nhonest}$, then $\localcomp\geq \compbound$. For $\replfact = {\nmalicious + \nhonest}$ and $\localcomp = \compbound$, we give a lower bound on $\commoh$. We then construct an \BGC scheme that requires $\localcomp \leq \compbound$ local computations, $\replfact = {\nmalicious+\nhonest}$, $\nround \leq \left(\nmalicious+1-\nhonest\right) \left\lceil \log_2\left(\frac{\ngrad}{\ngroup}\right) \right\rceil$ rounds and  $\commoh \leq \left(\nmalicious+1-\nhonest\right) \left( 2 \left\lceil \log_2\left( \frac{\ngrad}{\ngroup}\right) \right\rceil+ \frac{\nmalicious+3\nhonest}{2\log_2{\card{\galpha}}} \right)$.
\begin{figure}[t]
    \centering
    \resizebox{0.85\linewidth}{!}{
    \begin{tikzpicture}
\begin{axis}[
legend cell align={left}, 
legend columns={1}, 
legend style={color={rgb,1:red,0.0;green,0.0;blue,0.0}, draw opacity={1.0}, line width={1}, solid, fill={rgb,1:red,1.0;green,1.0;blue,1.0}, fill opacity={1.0}, text opacity={1.0}, font={{\fontsize{8 pt}{10.4 pt}\selectfont}}, text={rgb,1:red,0.0;green,0.0;blue,0.0}, cells={anchor={center}}, at={(0.98, 0.98)}, anchor={north east}}, 
axis background/.style={fill={rgb,1:red,1.0;green,1.0;blue,1.0}, opacity={1.0}}, 
anchor={north west}, 
width={93mm}, 
height={50mm}, 
xlabel={local computations $c$}, 
x tick style={color={rgb,1:red,0.0;green,0.0;blue,0.0}, opacity={1.0}}, 
x tick label style={color={rgb,1:red,0.0;green,0.0;blue,0.0}, opacity={1.0}, rotate={0}}, 
xmin={-0.8}, 
xmax={10.5}, 
xtick={{10,5,3,2,2,1,1,1,1,1,0}}, 
xticklabels={{$10$,$5$,$3$,$2$,$2$,$1$,$1$,$1$,$1$,$1$,$0$\\(DRACO)}}, 
xtick align={inside}, 
xticklabel style={align=center},
axis x line*={left}, 
ylabel={total worker to main node\\communication [\si{\giga\byte}]}, 
ylabel style={align=center},
ymajorgrids={true}, 
ymin={0},
ymax={402.890145778656}, 
axis y line*={left}, 
y axis line style={color={rgb,1:red,0.0;green,0.0;blue,0.0}, draw opacity={1.0}, line width={1}, solid}, 
colorbar={false}]
    \addplot[color={rgb,1:red,0.0;green,0.0;blue,0.0}, name path={32f6c51a-9b4d-4fbc-8a2f-b9530c7ca1b0}, area legend, fill={rgb,1:red,0.0667;green,0.4392;blue,0.6667}, fill opacity={1.0}, draw opacity={1.0}, line width={1}, solid]
        table[row sep={\\}]
        {
            \\
            9.6  204.89096694160253  \\
            9.6  0.0  \\
            10.4  0.0  \\
            10.4  204.89096694160253  \\
            9.6  204.89096694160253  \\
        }
        ;
    \addplot[color={rgb,1:red,0.0;green,0.0;blue,0.0}, name path={32f6c51a-9b4d-4fbc-8a2f-b9530c7ca1b0}, area legend, fill={rgb,1:red,0.0667;green,0.4392;blue,0.6667}, fill opacity={1.0}, draw opacity={1.0}, line width={1}, solid, forget plot]
        table[row sep={\\}]
        {
            \\
            4.6  223.51741838199086  \\
            4.6  0.0  \\
            5.4  0.0  \\
            5.4  223.51741838199086  \\
            4.6  223.51741838199086  \\
        }
        ;
    \addplot[color={rgb,1:red,0.0;green,0.0;blue,0.0}, name path={32f6c51a-9b4d-4fbc-8a2f-b9530c7ca1b0}, area legend, fill={rgb,1:red,0.0667;green,0.4392;blue,0.6667}, fill opacity={1.0}, draw opacity={1.0}, line width={1}, solid, forget plot]
        table[row sep={\\}]
        {
            \\
            2.6  242.143869822224  \\
            2.6  0.0  \\
            3.4  0.0  \\
            3.4  242.143869822224  \\
            2.6  242.143869822224  \\
        }
        ;
    \addplot[color={rgb,1:red,0.0;green,0.0;blue,0.0}, name path={32f6c51a-9b4d-4fbc-8a2f-b9530c7ca1b0}, area legend, fill={rgb,1:red,0.0667;green,0.4392;blue,0.6667}, fill opacity={1.0}, draw opacity={1.0}, line width={1}, solid, forget plot]
        table[row sep={\\}]
        {
            \\
            1.6  260.7703212623019  \\
            1.6  0.0  \\
            2.4  0.0  \\
            2.4  260.7703212623019  \\
            1.6  260.7703212623019  \\
        }
        ;
    \addplot[color={rgb,1:red,0.0;green,0.0;blue,0.0}, name path={32f6c51a-9b4d-4fbc-8a2f-b9530c7ca1b0}, area legend, fill={rgb,1:red,0.0667;green,0.4392;blue,0.6667}, fill opacity={1.0}, draw opacity={1.0}, line width={1}, solid, forget plot]
        table[row sep={\\}]
        {
            \\
            1.6  279.39677270222455  \\
            1.6  0.0  \\
            2.4  0.0  \\
            2.4  279.39677270222455  \\
            1.6  279.39677270222455  \\
        }
        ;
    \addplot[color={rgb,1:red,0.0;green,0.0;blue,0.0}, name path={32f6c51a-9b4d-4fbc-8a2f-b9530c7ca1b0}, area legend, fill={rgb,1:red,0.0667;green,0.4392;blue,0.6667}, fill opacity={1.0}, draw opacity={1.0}, line width={1}, solid, forget plot]
        table[row sep={\\}]
        {
            \\
            0.6  298.023224141992  \\
            0.6  0.0  \\
            1.4  0.0  \\
            1.4  298.023224141992  \\
            0.6  298.023224141992  \\
        }
        ;
    \addplot[color={rgb,1:red,0.0;green,0.0;blue,0.0}, name path={32f6c51a-9b4d-4fbc-8a2f-b9530c7ca1b0}, area legend, fill={rgb,1:red,0.0667;green,0.4392;blue,0.6667}, fill opacity={1.0}, draw opacity={1.0}, line width={1}, solid, forget plot]
        table[row sep={\\}]
        {
            \\
            0.6  316.64967558160424  \\
            0.6  0.0  \\
            1.4  0.0  \\
            1.4  316.64967558160424  \\
            0.6  316.64967558160424  \\
        }
        ;
    \addplot[color={rgb,1:red,0.0;green,0.0;blue,0.0}, name path={32f6c51a-9b4d-4fbc-8a2f-b9530c7ca1b0}, area legend, fill={rgb,1:red,0.0667;green,0.4392;blue,0.6667}, fill opacity={1.0}, draw opacity={1.0}, line width={1}, solid, forget plot]
        table[row sep={\\}]
        {
            \\
            0.6  335.27612702106126  \\
            0.6  0.0  \\
            1.4  0.0  \\
            1.4  335.27612702106126  \\
            0.6  335.27612702106126  \\
        }
        ;
    \addplot[color={rgb,1:red,0.0;green,0.0;blue,0.0}, name path={32f6c51a-9b4d-4fbc-8a2f-b9530c7ca1b0}, area legend, fill={rgb,1:red,0.0667;green,0.4392;blue,0.6667}, fill opacity={1.0}, draw opacity={1.0}, line width={1}, solid, forget plot]
        table[row sep={\\}]
        {
            \\
            0.6  353.90257846036303  \\
            0.6  0.0  \\
            1.4  0.0  \\
            1.4  353.90257846036303  \\
            0.6  353.90257846036303  \\
        }
        ;
    \addplot[color={rgb,1:red,0.0;green,0.0;blue,0.0}, name path={32f6c51a-9b4d-4fbc-8a2f-b9530c7ca1b0}, area legend, fill={rgb,1:red,0.0667;green,0.4392;blue,0.6667}, fill opacity={1.0}, draw opacity={1.0}, line width={1}, solid, forget plot]
        table[row sep={\\}]
        {
            \\
            0.6  372.5290298995096  \\
            0.6  0.0  \\
            1.4  0.0  \\
            1.4  372.5290298995096  \\
            0.6  372.5290298995096  \\
        }
        ;
    \addplot[color={rgb,1:red,0.0;green,0.0;blue,0.0}, name path={32f6c51a-9b4d-4fbc-8a2f-b9530c7ca1b0}, area legend, fill={rgb,1:red,0.0667;green,0.4392;blue,0.6667}, fill opacity={1.0}, draw opacity={1.0}, line width={1}, solid, forget plot]
        table[row sep={\\}]
        {
            \\
            -0.4  391.155481338501  \\
            -0.4  0.0  \\
            0.4  0.0  \\
            0.4  391.155481338501  \\
            -0.4  391.155481338501  \\
        }
        ;
    \addplot[color={rgb,1:red,0.0667;green,0.4392;blue,0.6667}, name path={9bed8f1c-8a16-44ab-bc3c-0b56859e86c4}, only marks, draw opacity={1.0}, line width={0}, solid, mark={*}, mark size={0.0 pt}, mark repeat={1}, mark options={color={rgb,1:red,0.0;green,0.0;blue,0.0}, draw opacity={0.0}, fill={rgb,1:red,0.0667;green,0.4392;blue,0.6667}, fill opacity={0.0}, line width={0.75}, rotate={0}, solid}, forget plot]
        table[row sep={\\}]
        {
            \\
            10.0  204.89096694160253  \\
            5.0  223.51741838199086  \\
            3.0  242.143869822224  \\
            2.0  260.7703212623019  \\
            2.0  279.39677270222455  \\
            1.0  298.023224141992  \\
            1.0  316.64967558160424  \\
            1.0  335.27612702106126  \\
            1.0  353.90257846036303  \\
            1.0  372.5290298995096  \\
            0.0  391.155481338501  \\
        }
        ;
\end{axis}
\end{tikzpicture}
    }
    \vspace{-1.1em}
    \caption{Tradeoff between total worker to main-node communication and the number of local computations for our scheme. The parameters are $\nmalicious = 10$, $1 \leq \nhonest \leq 11$, $\ngroup=1$, $\ngrad = \num{1e4}$, $\graddim=\num{1e6}$, $|\galpha|=2^{16}$. When considering total communication, the cost of the protocol is outweighed by the cost of transmitting the gradient values. The gap to our bound, given in \cref{thm:converse_T}, is less than \SI{5}{\kilo\byte}.
    For $\localcomp=0$ local computations, our scheme is equivalent to DRACO~\cite{chenDRACOByzantineresilientDistributed2018}.
    By requiring fewer workers, our scheme with $\nhonest=1$ reduces communication by \SI{48}{\percent} at the expense of at most $\localcomp=10$ local gradient computations at the \master. For comparison, each worker node performs $\ngrad=\num{1e4}$ gradient computations.}
    \label{fig:comm_vs_localcomp}
    \vspace{-1.4em}
\end{figure}

As depicted in \cref{fig:comm_vs_localcomp}, for realistic parameter ranges, the local computations drastically reduce the required communication. The communication overhead of the protocol $\commoh$ is outweighed by the initial transmission of $\iresp_{0, \wind}$. For a comparison between our achievability result and converse for $\commoh$, cf. \ARXIVonly{\cref{app:achievability_vs_converse}}\ISITonly{\cite[Appendix F]{hofmeisterBGC}}. 

Our lower bounds consider \Revision{an adversary that}\RevisionRemove{the case where the adversary} chooses the following behavior for the malicious workers.

\subsection{Symmetrization Attack for Bounds}\label{sec:attack}
    \Revision{We base parts of our proofs on the concept of a symmetrization attack, as explained in the following. The core idea is for the adversary to choose errors such that the \master cannot distinguish between different cases.}
    The adversary chooses a (potentially corrupted) value for each malicious worker and each partial gradient. 
    We denote worker $\worker$'s \emph{claimed} partial gradient results for $\pgrad$ as $\wpgrad{\gind}{\wind}\in\galpha^\graddim$ for all $\wind \in \range{\nworker}$ and $\gind\in\range{\ngrad}$.
    For honest workers we say $\wpgrad{\gind}{\wind} = \pgrad[\gind]$.
    Each worker $\worker$ computes their responses consistently based on those values, i.e.,
    \begin{align*}
        \noisyiresp_{\roundind, \wind} &= \encfun[\wind,\encind_{\wind,\roundind}]\left( \wpgrad{1}{\wind},\dots,\wpgrad{\ngrad}{\wind} \right),\\ 
        &\forall \roundind\in \{0, \dots, \nround\}, \wind \in \range{\nworker}, \encind_\wind \in \range{\nencfun}.
    \end{align*}
    For clarity of exposition, we first lay out how the adversary chooses the claimed gradient values for $\ngroup=1$ groups of size $\nmalicious+1$, i.e., for $\nhonest=1$, before we generalize to arbitrary $\ngroup, \nhonest \in \N$.

    For $\nhonest=1$ and $\ngroup=1$, the adversary draws a set $\disagreegradset \subseteq \range{\ngrad}$ of size $|\disagreegradset|=\lfloor \frac{\nmalicious}{\nhonest} \rfloor = \nmalicious$ uniformly at random and assigns each malicious worker $\worker$ a \Revision{unique} gradient index from $\disagreegradset$. \Revision{The adversary will introduce errors only for gradients $\pgrad[\tilde{\gind}]$, $\tilde{\gind} \in \disagreegradset$ and only for the one malicious worker that got assigned gradient index $\tilde{\gind}$.}
    \RevisionRemove{For each index, it sets the corresponding worker's claimed value to a distinct random vector unequal the true value and all other malicious worker's values to the true partial gradient value.}
    With probability $\frac{1}{2}$, it picks a single gradient index from $\disagreegradset$ uniformly at random and sets all malicious workers' values to the same erroneous random partial gradient value.
    The resulting claimed gradients take on the form as depicted in \cref{tab:key_error_pattern}.
\begin{table}[h]
    \vspace{-0.3cm}
    \begin{tabularx}{\columnwidth}{c|ccccccc}
        & $\wpgrad{1}{\wind}$ & $\wpgrad{2}{\wind}$ 
        & $\dots$  & $\wpgrad{\nmalicious}{\wind}$ & $\wpgrad{\nmalicious+1}{\wind}$ & $\dots$ & $\wpgrad{\ngrad}{\wind}$ \\ 
        \toprule
        $\worker[1]$ & $\pgrad[1]^{\prime\prime}$ & $\pgrad[2]^\prime$ & $\dots$  & $\pgrad[\nmalicious]^\prime$ & $\pgrad[\nmalicious+1]^\prime$ & $\dots$ & $\pgrad[\ngrad]^\prime$ \\ 
        $\worker[2]$ & $\pgrad[1]^\prime$ & $\pgrad[2]^{\prime\prime}$ & $\dots$  & $\pgrad[\nmalicious]^\prime$ & $\pgrad[\nmalicious+1]^\prime$ & $\dots$ & $\pgrad[\ngrad]^\prime$ \\ 
        $\vdots$ & $\vdots$ & $\vdots$ & $\ddots$ & $\vdots$ & $\vdots$ & $\dots$ & $\vdots$ \\
        $\worker[\nmalicious]$ & $\pgrad[1]^\prime$ & $\pgrad[2]^\prime$ & $\dots$ & $\pgrad[\nmalicious]^{\prime\prime}$ & $\pgrad[\nmalicious+1]^\prime$ & $\dots$ & $\pgrad[\ngrad]^\prime$ \\ 
       $\worker[\nmalicious+1]$ & $\pgrad[1]^\prime$ & $\pgrad[2]^\prime$ & $\dots$  & $\pgrad[\nmalicious]^\prime$ & $\pgrad[\nmalicious + 1]^\prime$ & $\dots$ & $\pgrad[\ngrad]^\prime$ \\ 
    \end{tabularx}
    \caption{claimed partial gradients for symmetrization attack.}
    \vspace{-0.5cm}
    \label{tab:key_error_pattern}
\end{table}
    For partial gradients with index in $\disagreegradset$, there are two competing values $\pgrad^{\prime}$ and $\pgrad^{\prime\prime}$ whereas for all other gradient indices the claimed values by all workers agree.

    For $\nhonest > 1$, the adversary randomly partitions the malicious workers into $\lfloor \nmalicious / \nhonest \rfloor -1$ sets of size $\nhonest$ and (depending on divisibility) one group of size $\nmalicious \mod \nhonest$.
    Each of the sets of size $\nhonest$ behaves like one malicious worker in the case $\nhonest=1$.
    The workers in the remaining set of less than $\nhonest$ workers (if non-empty) pick their claimed gradients randomly either like a random other set of workers or like the honest workers.
    The resulting claimed gradients take on the form as depicted in \ISITonly{\cite[Appendix B]{hofmeisterBGC}}\ARXIVonly{\cref{app:large_table}}.

    For $\ngroup > 1$, the adversary \RevisionRemove{chooses the first group and }chooses the claimed gradients \Revision{in the first group} according to the attack strategy for $\ngroup^\prime=1$ and $\ngrad^\prime=\ngrad / \ngroup$\RevisionRemove{ in that group}. In all other groups, the claimed gradients \Revision{equal}\RevisionRemove{are equal to} their true values.

\subsection{Fundamental Limits}

\begin{theorem}[Lower bound on $\localcomp$ and $\replfact$]
Suppose that $\nworker = \ngroup (\nmalicious + \nhonest)$ for integers $\ngroup,\nhonest \geq 1$. For any tuple (\allocmat, \encfunset, \decfun, \proto), with $\allocmat$ as in \eqref{eq:fractional_repetition}, to be a (\nround,\localcomp,\replfact,\commoh)-\BGC scheme, it holds that if $\replfact \leq \nmalicious + \nhonest$, then $\localcomp\geq \compbound$ (conversely,  if $\localcomp < \compbound$ then $\replfact > {\nmalicious + \nhonest}$) for any number of rounds $\nround$ and any communication overhead $\commoh$. 
\label{thm:converse_C}
\end{theorem}
\begin{IEEEproof}
    We demonstrate that any tuple $(\allocmat, \encfunset, \decfun, \proto)$ with fractional repetition data allocation which uses $\localcomp \leq \compbound - 1$ local computations cannot be an \BGC scheme. Specifically, we show that the malicious workers can always perform a symmetrization attack preventing the main node from deterministically computing the full gradient. 
    Let the malicious workers behave as explained in \cref{sec:attack}. 

    We abstract $\encfunset$ and $\proto_1$ by assuming that the values of all partial gradients $\wpgrad{\gind}{\wind}$ computed by the workers are available at the \master. Note that regardless of $\encfunset$ and $\proto_1$ the \master cannot gain any additional information from the workers' responses.
    We now show that for $\localcomp \leq \compbound-1$ there exists no choice of a decoding function $\decfun$ and $\proto_2$ for which the \master deterministically outputs the true full gradient. To that end, for any possible $\proto_2$, we give two cases for which the inputs to the decoding function are identical but the true full gradients differ.

  The claimed partial gradients take on values as in \cref{tab:key_error_pattern} for $\nhonest=1$ and \ISITonly{\cite[Appendix B]{hofmeisterBGC}}\ARXIVonly{\cref{app:large_table}} for $\nhonest>1$. There are $\compbound$ partial gradients, say $\pgrad[1],\ldots,\pgrad[\compbound]$, such that for each partial gradient $\pgrad[\gind]$ worker $\worker[\gind]$ sends a value $\wpgrad{\gind}{\gind} = \pgrad[\gind]^{\prime\prime}$ that is different from the value $\pgrad[i]^\prime$ sent by all other workers.
For any list $\lgindset^{(\nround)}$ of locally computed gradients of size $|\lgindset^{(\nround)}| \leq \nmalicious-1$ (produced by any $\proto_2$) there exists an index $\widetilde{\gind} \in [\nmalicious]$ such that $\widetilde{\gind} \notin \lgindset^{(\nround)}$.
Consider \RevisionRemove{the following}\Revision{these} two cases:\begin{enumerate}[label=Case \arabic{*}:, wide=1.5\parindent, leftmargin=*]
    \item $\pgrad[\gind] = \pgrad[\gind]^\prime \;\forall \gind \in [\frac{\ngrad}{\ngroup}]$ and 
    \item $\pgrad[\gind] = \pgrad[\gind]^\prime \;\forall \gind \in [\frac{\ngrad}{\ngroup}] \setminus \{\widetilde{\gind}\}$ and $\pgrad[\widetilde{\gind}] = \pgrad[\widetilde{\gind}]^{\prime\prime}$.
\end{enumerate}
It is easy to see that both cases occur with non-zero probability according to the attack strategy in \cref{sec:attack}. In both cases, the inputs to $\decfun$ only depend on the fixed $\lgindset^{(\nround)}$, the $\wpgrad{\gind}{\wind},\;\wind \in [\nmalicious +1],\;\gind\in[\frac{\ngrad}{\ngroup}]$ and the $\pgrad[\gind],\; \gind \in \lgindset^{(\nround)}$, all of which take on identical values in both cases.
The value of the full gradient $\tgrad$, however, is $\sum_{\gind \in [\ngrad]} \pgrad[\gind]^\prime$ in Case 1 and $\pgrad[\widetilde{\gind}]^{\prime\prime} + \sum_{\gind \in [\ngrad] \setminus \{\widetilde{\gind}\}} \pgrad[\gind]^\prime$ in Case 2. Hence, for $\localcomp < \compbound$ no decoding function can deterministically produce the correct full gradient.
\end{IEEEproof}

\begin{theorem}[Lower bound on $\commoh$ for fixed $\localcomp$ and $\replfact$]
Suppose that $\nworker = \ngroup (\nmalicious + \nhonest)$ for integers $\ngroup,\nhonest \geq 1$. For any tuple (\allocmat, \encfunset, \decfun, \proto), with $\allocmat$ as in \eqref{eq:fractional_repetition}, to be a (\nround,\localcomp,\replfact,\commoh)-\BGC scheme with $\replfact = {\nmalicious + \nhonest}$ and $\localcomp = \compbound$, then it must hold that
  \vspace{-1em}
    \begin{align*}
        \commoh \geq {\log_{\alphasize} \binom{\ngrad/\ngroup}{\lfloor \nmalicious / \nhonest \rfloor}}.%
    \end{align*}
  \vspace{-0.6em}
    \label{thm:converse_T}
\end{theorem}

\vspace{-0.3cm}
\begin{IEEEproof}
  The proof is given in~\ISITonly{\cite[Appendix D]{hofmeisterBGC}}\ARXIVonly{\cref{app:proof_lower_bound_kappa_rho}}.
\end{IEEEproof}

\subsection{Construction of an \BGC Scheme}

\begin{theorem}
  The scheme constructed below is an \BGC scheme with a parameter $\nhonest$ such that $1\leq \nhonest \leq \nmalicious+1$ and requires $\nround \leq \left(\nmalicious+1-\nhonest\right) \left\lceil \log_2\left(\frac{\ngrad}{\ngroup}\right) \right\rceil$ rounds, $\localcomp \leq \compbound$ local gradient computations, $\replfact = {\nmalicious+\nhonest}$, and $\commoh \leq \left(\nmalicious+1-\nhonest\right) \left( 2 \left\lceil \log_2\left( \frac{\ngrad}{\ngroup}\right) \right\rceil+ \frac{\nmalicious+3\nhonest}{2\log_2{\card{\galpha}}} \right)$. 
    \label{thm:scheme}
\end{theorem}
\begin{IEEEproof}
The proof is given in \ISITonly{\cite[Appendix E]{hofmeisterBGC}}\ARXIVonly{\cref{app:scheme_proof}}.
\end{IEEEproof}

We construct an $(\nround, \localcomp, \replfact, \commoh)$-\BGC scheme that has a replication factor $\replfact={\nmalicious+\nhonest}$, $\nhonest \geq 1$, and requires the optimal local computation load $\localcomp \leq \compbound$ at the \master. The protocol $\proto$ runs for $\nround \leq \left(\nmalicious+1-\nhonest\right) \left\lceil \log_2\left(\frac{\ngrad}{\ngroup}\right) \right\rceil$ rounds and achieves a communication overhead \mbox{$\commoh \leq \left(\nmalicious+1-\nhonest\right) \left( 2 \left\lceil \log_2\left( \frac{\ngrad}{\ngroup}\right) \right\rceil+ \frac{\nmalicious+3\nhonest}{2\log_2{\card{\galpha}}} \right)$}. Our scheme uses a fractional repetition data assignment with $\ngroup$ groups each of size $\nmalicious+\nhonest$. Informally, in each group, the \master runs an elimination tournament consisting of matches (similar to the one explained in~\cref{ex:toy_problem}) between pairs of workers that return contradicting responses. \RevisionRemove{An algorithmic description of the elimination tournament is given in
  Algorithm ...
for the general case.} For clarity of exposition, we explain the idea of our scheme for the special case of $\nhonest=1$. The general case for $\nhonest\geq 2$ follows similar steps \ISITonly{\cite[Appendix C]{hofmeisterBGC}}\ARXIVonly{\cref{app:general_case}}. %

The tournament consists of a series of matches between two workers. 
During a match, each worker constructs a binary tree based on their computed partial gradients, which we refer to as the match tree.
The root of the tree is labeled by the sum of all partial gradients computed at that worker ($\noisyiresp_{0,\wind}$).
The child nodes are constructed based on the partial gradients that are contained in the parent node.
Each node has two children: the first one is labeled by the sum of the first half of the parent node's partial gradients; the second one is labeled by the sum of the second half of the parent node's partial gradients.
Thus, the labeling is done such that the sum of the labels of any two siblings gives the label of their parent node.

Proceeding in this way recursively, each worker ends up with the leaves of the tree being labeled by individual partial gradients.
For example, when $\ngroup = 1$ and $\ngrad=4$, the tree is depicted in \cref{fig:binary_add_tree}.
During a match, the \master requests the labels for particular nodes in this tree from the two competing workers, and compares them. 
If the root labels of two match trees differ, then by definition there must be a child node for which the corresponding label differs in those trees. By induction, it is clear that there has to be a path from the root to a leaf, such that the corresponding labels of all involved nodes differ between the two match trees. Applying this observation to the example in \cref{fig:binary_add_tree}, if $\wpgrad{1}{\wind}+\wpgrad{2}{\wind}+\wpgrad{3}{\wind}+\wpgrad{4}{\wind}$ is different for two workers, then $\wpgrad{1}{\wind}+\wpgrad{2}{\wind}$ or $\wpgrad{3}{\wind}+\wpgrad{4}{\wind}$ must also differ (or both). In the latter case, we end up with $\wpgrad{3}{\wind}$ or $\wpgrad{4}{\wind}$ being different between the workers.
\begin{figure}[t]
  \vspace{-0.8em}
    \centering
    \resizebox{0.9\linewidth}{!}{
    \begin{tikzpicture}
\def\angle{45}
\node {$\wpgrad{1}{\wind}+\wpgrad{2}{\wind}+\wpgrad{3}{\wind}+\wpgrad{4}{\wind}$} [sibling distance = 5cm,level distance=0.8cm]
    child {node (a) {$\wpgrad{1}{\wind}+\wpgrad{2}{\wind}$} [sibling distance = 2.5cm,level distance=1cm]
        child {node {$\wpgrad{1}{\wind}$}}
        child {node (y) {$\wpgrad{2}{\wind}$}}
    } 
    child {node (d) {$\wpgrad{3}{\wind}+\wpgrad{4}{\wind}$} [sibling distance = 2.5cm,level distance=0.8cm]
        child {node (z) {$\wpgrad{3}{\wind}$}}
        child {node {$\wpgrad{4}{\wind}$}}
    };
\end{tikzpicture}
    }
  \vspace{-1em}
    \caption{Example of a match tree for $\worker$ and parameters $\ngroup=1,\ngrad=4$.}
    \label{fig:binary_add_tree}
    \vspace{-0.5cm}
\end{figure}

The match starts at the root. In this case, each worker $\worker$ would have already sent the node label in $\iresp_{0,\wind}$ to the \master, i.e.,
  \vspace{-0.6em}
\begin{equation*}
    \iresp_{0,\wind} = \encfun[\wind,1]\left( \pgrad[1],\dots,\pgrad[\ngrad] \right) = \sum_{\substack{ \gind \in \range{\ngrad} \\ \allocmat[\gind, \wind] = 1}} \pgrad[\gind].
  \vspace{-0.8em}
\end{equation*}
Note that, without errors, all workers' messages agree within a group.
In case of discrepancies between the responses $\noisyiresp_{0,\wind}$ of the workers within a group, the \master selects a pair of disagreeing workers $\worker[\wind_1],\worker[\wind_2]$ and further descends in the tree as follows.

For every node in the tree, the \master requests and compares the left child's label from both workers, encoded in $\iresp_{\roundind,\indone}$ and $\iresp_{\roundind,\indtwo}$. The \master then moves on to a child whose label the workers disagree on.
\RevisionRemove{Note that,} Based on the current node's label and the left child's label, the \master can infer the right child's label.
If the competing workers agree on the left child's label, they must disagree on the right child's label.
This procedure is repeated until a leaf is reached.
Note that, even if the workers send inconsistent labels at each round, this procedure is guaranteed to reach a leaf node for which the (sent or inferred) values of the individual gradient is different for the two workers. \RevisionRemove{Algorithm ...
formalizes this procedure for $\ngroup=1$.}

\RevisionRemove{Note that} It is possible to reduce the communication load by picking a coordinate $\compind$ in which the workers' initial responses disagree, i.e., $\comp{ \iresp_{0,\indone} } \neq \comp{ \iresp_{0,\indtwo} }$. That is, the workers only encode the $\compind$-th coordinate of the node labels. By this method, it is still guaranteed that the \master can identify a partial gradient for which the competing workers disagree.

Having identified disagreeing values $\wpgrad{\gind}{\wind_1}$ and $\wpgrad{\gind}{\wind_2}$ of a partial gradient $\pgrad[\gind]$, the \master computes the correct \Revision{value} of this partial gradient locally. It then marks the worker(s) whose values differ from the value computed locally as malicious.
The algorithm ends up with disagreeing leaf labels by design. \RevisionRemove{Therefore,} Each match is guaranteed to eliminate at least one malicious worker and no honest worker.
After performing at most $\nmalicious$ matches, the \master is guaranteed to identify all malicious workers.
The \master takes the encoded gradient of one worker identified as being honest from each group and sums up the group-wise results to obtain $\gestim=\tgrad$.

Algorithmic descriptions of the elimination tournament and a match between workers are in~\ISITonly{\cite[Appendix H]{hofmeisterBGC}}\ARXIVonly{\cref{app:algorithms}}.

\subsection{Discussion}
According to \cref{thm:converse_C}, the interactive protocol of our scheme achieves the lowest possible number of locally computed partial gradients for the fractional repetition data assignment and $\replfact = \nmalicious+\nhonest$. Note that for $\nhonest \geq \nmalicious+1$, i.e., $\replfact \geq 2\nmalicious+1$, no local computation is necessary. In fact, since there is only one set of consistent workers $\workerset$ that has $\card{\workerset} \geq \nhonest$ per fractional repetition group, the scheme immediately identifies $\workerset$ as the honest set and terminates without any additional computation or communication. Thus, as shown in~\cite{chenDRACOByzantineresilientDistributed2018}, \Revision{this is optimal.}\RevisionRemove{our scheme achieves the optimal performance in this case.} Although we consider $1 \leq \nhonest \leq \nmalicious+1$ in \cref{thm:scheme} for technical reasons, the scheme works for any $\nhonest \geq 1$. \Revision{We remark that since $\ngrad \gg \nmalicious$ in state-of-the-art machine learning deployments, the local computations cause only a relatively small load at the \master.}
Note that the achievable bound on $\commoh$ in \cref{thm:scheme} is off from the converse bound in \cref{thm:converse_T} by a constant factor asymptotically, see \ISITonly{\cite[Appendix F]{hofmeisterBGC}}\ARXIVonly{\cref{app:achievability_vs_converse}}. The reason is that we use a rather simple and conservative lower bound on the amount of information that is required to be transmitted by the workers in \cref{thm:converse_T}. Among others, we assume that all workers have knowledge about the malicious worker's identities and also the malicious workers contribute useful information. \RevisionRemove{Improvement of the} \Revision{A tighter} converse bound is left \RevisionRemove{for}\Revision{as} future work. \Revision{Finally, we remark that although the communication complexity of our scheme is quadratic in $\nmalicious$, this value is not very large in practice.}

\section{Conclusion}
We considered the problem of distributed learning in the presence of Byzantine computation errors. We introduced a framework that extends the known gradient coding framework by adding an interactive light communication between the \master and the workers and verifying local computations at the \master.
In the scope of this framework, we proposed a new scheme that can tolerate $\nmalicious$ malicious workers with a computational redundancy of $\nmalicious+\nhonest$ for any $\nhonest \geq 1$. We showed that with a fractional repetition data assignment, the scheme achieves the optimal number of local computations at the \master. Future work includes the improvements of the converse and achievability bounds, the generalization of the fundamental limits to a broader class of data assignments, and the investigation into optimal Byzantine-resilient gradient coding schemes.

\balance
\bibliographystyle{IEEEtran}
\bibliography{references}

\ARXIVonly{
  \clearpage
  \newpage

  \appendix
  \subsection{Elaborate Example on How to Catch Liars Efficiently}
  \label{app:biggerexample}
  The following example demonstrates the main problem studied in the paper and gives intuition for our main results.
 Consider a game among friends. \playA, \playB and \playC play against \playboss.
\begin{enumerate}
    \item \playA, \playB, and \playC secretly agree on a list of four integers $g_1, \dots, g_8$ and put them in separate envelopes.
    \item Two players between \playA, \playB, and \playC are designated as liars, without \playboss knowing which is which. The remaining player has to be truthful.
    \item \playboss's goal is to find the sum of the numbers $g_1 + \dots + g_8$. The game is played in rounds. In each round \playbosspronoun can first \begin{enumerate*}
        \item ask the other players questions about $g_1, \dots, g_8$ and then
        \item look in any number of envelopes.
    \end{enumerate*}
    \item \playboss needs to find the correct sum every time. 
\end{enumerate}
What is the minimum number of envelopes \playbosspronoun needs to look inside?
If \playbosspronoun checks the minimum number of envelopes, how many questions does \playbosspronoun need to ask? 

If the number of questions asked were unlimited, \playbosspronoun could ask each player for all values $g_1, \dots, g_8$. With $\exwpgrad{\gind}{\wind}$ we denote the value the value player $\wind\in \{\text{\playA}, \text{\playB}, \text{\playC}\}$ claims is in the envelope for $g_\gind$, $\gind \in [8]$.
Since the players only answer questions about these numbers, and assuming the liars are smart enough to avoid contradictions so as to not be detected, \playboss cannot gain any more information by asking more questions.

Not in every case can \playboss identify the correct sum based just on these answers.
For example, \cref{tab:example_key_error_pattern} shows three cases that are indistinguishable based on the values of all $\exwpgrad{\gind}{\wind}$.
\begin{table}[h]
    \begin{minipage}{0.45\columnwidth}
    \begin{tabularx}{\columnwidth}{l*{4}{X}}
        & $\exwpgrad{1}{\wind}$ & $\exwpgrad{2}{\wind}$ & \exwpgrad{3}{\wind} & \dots \\
        \toprule
        \playA & $1$ & $3$ & $4$ & \dots \\
        \playB & $2$ & $7$ & $4$ & \dots \\
        \playC & $2$ & $3$ & $4$ & \dots \\
    \end{tabularx}
    \end{minipage}%
    \hfill
    \begin{minipage}{0.49\columnwidth}
        \begin{align*}
            \text{\emph{Case 1:  }} 
                            & g_1=2;\ g_2=3 \\
                            & \text{\playA and \playB lie} \\
                \text{\emph{Case 2:  }} 
                            & g_1=1;\ g_2=3 \\
                            & \text{\playB and \playC lie} \\
                \text{\emph{Case 3:  }}
                            & g_1=2;\ g_2=7 \\
                            & \text{\playA and \playC lie}
        \end{align*}
    \end{minipage}
    \caption{Indistinguishable Cases.}
    \label{tab:example_key_error_pattern}
\end{table}

\playboss picks any $g_i$, $i\in[8]$ on which at least two players' answers disagree and checks the corresponding envelope. \playbossPronoun is guaranteed to identify at least one liar. After eliminating the identified liar(s) \playbosspronoun repeats the process.
After opening at most two envelopes it is guaranteed that all non-eliminated players, including the honest one, agree on the sum $g_1 + \dots + g_8$ and it is therefore correct.
It can be verified, that the cases in \cref{tab:example_key_error_pattern} cannot be distinguished after opening any one envelope. Thus, using this strategy \playboss opens the smallest possible number of envelopes in the worst case.

Now, \playboss additionally wishes to minimize the number of questions asked.
With the above strategy, \playboss asks each player eight questions, for a total of $24$.

\playboss realizes that using multiple rounds, \playbosspronoun can reduce the number of questions \playbosspronoun needs to ask using the following recursive procedure.
First, \playbosspronoun asks every player for the desired sum $g_1 + \dots + g_8$ and the sum of the first half of the values, $g_1 + g_2 + g_3 + g_4$.
\playboss picks any two players, whose values for the total sum differ.
\playbossPronoun infers their values for $g_5 + \dots + g_8$ as $(g_1 + \dots + g_8) - (g_1 + \dots + g_4)$.
If the two players agree on $g_1 + \dots + g_4$, they are guaranteed to disagree on $g_5 + \dots + g_8$.
Assume they disagree on $g_5 + \dots + g_8$, then \playboss requests $g_5 + g_6$ from both players.
Again, they are guaranteed to disagree either on $g_5 + g_6$ or $g_7 + g_8$.
Assume they disagree on $g_5+g_6$. \playboss asks for $g_5$.
After having obtained a single integer, either $g_5$ or $g_6$ on which the two players disagree, \playboss opens the corresponding envelope and thus identifies and eliminates at least one liar.
In the worst case \playboss has to repeat the above procedure once more on the remaining two values.
Now, \playbosspronoun asks only $15$ questions as opposed to the $24$ from before.

  \subsection{Symmetrization Attack Table}
  \label{app:large_table}
  \begin{table}[h]
  \resizebox{\columnwidth}{!}{%
    \begin{tabularx}{1.1\columnwidth}{l|lllllll}
        & $\wpgrad{1}{\wind}$ & $\wpgrad{2}{\wind}$ 
        & $\dots$  & $\wpgrad{\compbound}{\wind}$ & $\wpgrad{\compbound+1}{\wind}$ & $\dots$ & $\wpgrad{\ngrad}{\wind}$ \\ 
        \addlinespace[3pt]
        \toprule
        $\worker[1]$ & $\pgrad[1]^{\prime\prime}$ & $\pgrad[2]^\prime$ & $\dots$  & $\pgrad[\lfloor \frac{\nmalicious}{\nhonest} \rfloor]^\prime$ & $\pgrad[\lfloor \frac{\nmalicious}{\nhonest} \rfloor + 1]^\prime$ & $\dots$ & $\pgrad[\ngrad]^\prime$ \\ 
        $\worker[2]$ & $\pgrad[1]^{\prime\prime}$ & $\pgrad[2]^\prime$ & $\dots$  & $\pgrad[\lfloor \frac{\nmalicious}{\nhonest} \rfloor]^\prime$ & $\pgrad[\lfloor \frac{\nmalicious}{\nhonest} \rfloor + 1]^\prime$ & $\dots$ & $\pgrad[\ngrad]^\prime$ \\ 
        $\vdots$ & $\vdots$ & $\vdots$ & $\vdots$ & $\vdots$ & $\vdots$ & $\dots$ & $\vdots$ \\
        $\worker[\nhonest]$ & $\pgrad[1]^{\prime\prime}$ & $\pgrad[2]^\prime$ & $\dots$  & $\pgrad[\lfloor \frac{\nmalicious}{\nhonest} \rfloor]^\prime$ & $\pgrad[\lfloor \frac{\nmalicious}{\nhonest} \rfloor + 1]^\prime$ & $\dots$ & $\pgrad[\ngrad]^\prime$ \\ 
        \addlinespace[3pt]
        \hdashline
        \addlinespace[3pt]
        $\worker[\nhonest+1]$ & $\pgrad[1]^\prime$ & $\pgrad[2]^{\prime\prime}$ & $\dots$  & $\pgrad[\lfloor \frac{\nmalicious}{\nhonest} \rfloor]^\prime$ & $\pgrad[\lfloor \frac{\nmalicious}{\nhonest} \rfloor + 1]^\prime$ & $\dots$ & $\pgrad[\ngrad]^\prime$ \\ 
        $\worker[\nhonest+2]$ & $\pgrad[1]^\prime$ & $\pgrad[2]^{\prime\prime}$ & $\dots$  & $\pgrad[\lfloor \frac{\nmalicious}{\nhonest} \rfloor]^\prime$ & $\pgrad[\lfloor \frac{\nmalicious}{\nhonest} \rfloor + 1]^\prime$ & $\dots$ & $\pgrad[\ngrad]^\prime$ \\ 
        $\vdots$ & $\vdots$ & $\vdots$ & $\vdots$ & $\vdots$ & $\vdots$ & $\dots$ & $\vdots$ \\
        $\worker[2\nhonest]$ & $\pgrad[1]^\prime$ & $\pgrad[2]^{\prime\prime}$ & $\dots$  & $\pgrad[\lfloor \frac{\nmalicious}{\nhonest} \rfloor]^\prime$ & $\pgrad[\lfloor \frac{\nmalicious}{\nhonest} \rfloor + 1]^\prime$ & $\dots$ & $\pgrad[\ngrad]^\prime$ \\ 
        \addlinespace[3pt]
        \hdashline
        \addlinespace[3pt]
        $\vdots$ & $\vdots$ & $\vdots$ & $\ddots$ & $\vdots$ & $\vdots$ & $\dots$ & $\vdots$ \\
        \addlinespace[3pt]
        \hdashline
        \addlinespace[3pt]
        $\worker[\lfloor \frac{\nmalicious}{\nhonest} \rfloor \nhonest - \nhonest +1]$ & $\pgrad[1]^\prime$ & $\pgrad[2]^\prime$ & $\dots$ & $\pgrad[\lfloor \frac{\nmalicious}{\nhonest} \rfloor]^{\prime\prime}$ & $\pgrad[\lfloor \frac{\nmalicious}{\nhonest} \rfloor + 1]^\prime$ & $\dots$ & $\pgrad[\ngrad]^\prime$ \\ 
        $\worker[\lfloor \frac{\nmalicious}{\nhonest} \rfloor \nhonest - \nhonest +2]$ & $\pgrad[1]^\prime$ & $\pgrad[2]^\prime$ & $\dots$ & $\pgrad[\lfloor \frac{\nmalicious}{\nhonest} \rfloor]^{\prime\prime}$ & $\pgrad[\lfloor \frac{\nmalicious}{\nhonest} \rfloor + 1]^\prime$ & $\dots$ & $\pgrad[\ngrad]^\prime$ \\ 
        $\vdots$ & $\vdots$ & $\vdots$ & $\vdots$ & $\vdots$ & $\vdots$ & $\dots$ & $\vdots$ \\
        $\worker[\lfloor \frac{\nmalicious}{\nhonest} \rfloor\nhonest]$ & $\pgrad[1]^\prime$ & $\pgrad[2]^\prime$ & $\dots$ & $\pgrad[\lfloor \frac{\nmalicious}{\nhonest} \rfloor]^{\prime\prime}$ & $\pgrad[\lfloor \frac{\nmalicious}{\nhonest} \rfloor + 1]^\prime$ & $\dots$ & $\pgrad[\ngrad]^\prime$ \\ 
        \addlinespace[3pt]
        \hdashline
        \addlinespace[3pt]
        $\worker[\lfloor \frac{\nmalicious}{\nhonest} \rfloor\nhonest +1]$ & $\pgrad[1]^\prime$ & $\pgrad[2]^\prime$ & $\dots$ & $\pgrad[\lfloor \frac{\nmalicious}{\nhonest} \rfloor]^{\prime}$ & $\pgrad[\lfloor \frac{\nmalicious}{\nhonest} \rfloor + 1]^\prime$ & $\dots$ & $\pgrad[\ngrad]^\prime$ \\ 
        $\worker[\lfloor \frac{\nmalicious}{\nhonest} \rfloor\nhonest +2]$ & $\pgrad[1]^\prime$ & $\pgrad[2]^\prime$ & $\dots$ & $\pgrad[\lfloor \frac{\nmalicious}{\nhonest} \rfloor]^{\prime}$ & $\pgrad[\lfloor \frac{\nmalicious}{\nhonest} \rfloor + 1]^\prime$ & $\dots$ & $\pgrad[\ngrad]^\prime$ \\ 
        $\vdots$ & $\vdots$ & $\vdots$ & $\vdots$ & $\vdots$ & $\vdots$ & $\dots$ & $\vdots$ \\
        $\worker[\nmalicious+\nhonest]$ & $\pgrad[1]^\prime$ & $\pgrad[2]^\prime$ & $\dots$  & $\pgrad[\lfloor \frac{\nmalicious}{\nhonest} \rfloor]^\prime$ & $\pgrad[\lfloor \frac{\nmalicious}{\nhonest} \rfloor + 1]^\prime$ & $\dots$ & $\pgrad[\ngrad]^\prime$ \\ 
    \end{tabularx}
  }
  \vspace{1em}
    \caption{Claimed Partial Gradients for Symmetrization Attack.}
    \label{tab:key_error_pattern_generalized}
\end{table}

  \FloatBarrier
  \subsection{General Scheme Description}\label{app:general_case}
  In this section, we explain the extension of our scheme for the general case of $\nhonest\geq 1$. In this case, we also exploit the fact that there are $\nhonest$ honest workers that are guaranteed to agree in their responses. Therefore, the \master can eliminate all responses that are supported by only less than $\nhonest$ workers. Furthermore, a response must be a correct response if it is supported by more than $\nmalicious$ workers. Therefore, the \master does only need to consider sets $\workerset \in \range{\nworker}$ of workers that agree on their responses, and which satisfies $\nhonest \leq \card{\workerset} \leq \nmalicious$. For the clarity of presentation, we thus, only consider those sets as inputs to \cref{alg:elimination_tournament}.

Additionally, we leverage a little more communication to reduce the number of local gradient computations to at most $\compbound$. The intuition here is that the \master can pick a representative from every set $\workerset$ and run matches between the representatives. Note however, that even if the \master identifies a representative of $\workerset$ as malicious by a local computation, this does not imply that every worker in $\workerset$ is malicious. For example, a malicious worker could return the same initial response as the honest workers, but when being picked as a representative, intentionally send wrong node labels. We overcome this by asking all workers in a set $\workerset$ to either commit or not commit to the representative's leaf label, which has been identified in a match. In order to force the \master to run a local computation, there must be at least $\nhonest$ workers in $\workerset$ that commit to the representative's leaf label. If not, we can mark all the workers that committed to the label as malicious immediately. If there are at least $\nhonest$ commitments to a malicious leaf label, the \master can mark all those workers (at least $\nhonest$) as malicious by running a local computation. This procedure is reflected by the elimination tournament in \cref{alg:elimination_tournament}.

  \FloatBarrier
  \subsection{Proof of \cref{thm:converse_T}}\label{app:proof_lower_bound_kappa_rho}
      We consider a single group consisting of $\nworker = \nmalicious + \nhonest$ workers. Since the datasets per group, as well as the sets of workers, are disjoint, the communication necessary for $\ngroup$ groups of size $\ngrad/\ngroup$ is at least as big as for $\ngroup^\prime=1$ group of size $\ngrad/\ngroup$.
    Further, we assume the behavior of the malicious workers as in \cref{sec:attack}.

    To proceed with the proof, we need the following direct consequence of the proof technique used to prove \cref{thm:converse_C}.

\begin{corollary}[Computation of Disagreement Gradients]\label{cor:disagreement}
    For any (\nround,\localcomp,\replfact,\commoh)-\BGC scheme (\allocmat, \encfunset, \decfun, \proto), with $\replfact = \nmalicious + \nhonest$ and \allocmat\ as in \eqref{eq:fractional_repetition},
    if the adversary behaves according to \cref{sec:attack}, then the list of locally computed gradients must contain the list of all gradients on which the workers disagree, i.e.,
  $\forall \widetilde{\gind} \in \disagreegradset:  \widetilde{\gind} \in \lgindset^{(\nround)}$.%
    \label{cor:comp_disag_grads}
\end{corollary}
\begin{IEEEproof}
    The proof follows the same steps as for \cref{thm:converse_C}. Note that if there exists an index $\widetilde{\gind} \in \disagreegradset: \widetilde{\gind} \notin \lgindset^{(\nround)}$, there are two cases that cannot be distinguished based on the information available at the \master.
\end{IEEEproof}

    According to \cref{cor:comp_disag_grads}, for $\localcomp = \lfloor \frac{\nmalicious}{\nhonest} \rfloor$, we require every item of $\disagreegradset$ to be in $\lgindset^{(T)}$, while $\localcomp = |\lgindset^{(\nround)}| = |\disagreegradset| = \compbound$.
    In other words, the \master must exactly compute all the gradients in $\disagreegradset$ locally. If there is a non-zero probability that a different gradient is computed locally, the scheme cannot be a valid \BGC scheme.
    Overall, $\disagreegradset$ must be uniquely determined by the \master's available information at the end of the protocol, leading to
        $\En(\disagreegradset \given \wtmdata, \locgradset) = 0$,
    where $\wtmdata$ denotes the list of random variables corresponding to all data transmitted from all workers to the \master and \locgradset denotes the list of the random variables corresponding to the values of the locally computed gradients.

    Using this we have
    \begin{align}
        \En(\wtmdata \given \locgradset) 
                    \label{eq:defmi}
                    &\geq \I(\wtmdata; \disagreegradset \given \locgradset) \\ 
                    \notag
                    &= \En(\disagreegradset \given \locgradset) - \En(\disagreegradset \given \locgradset, \wtmdata) \\
                    \notag
                    &= \En(\disagreegradset \given \locgradset) \\
                    \label{eq:indep_disagreegradset}
                    &= \En(\disagreegradset) \\
                    &= \log_\alphasize \binom{\ngrad / \ngroup}{\lfloor \nmalicious / \nhonest \rfloor},\label{eq:uniform}
\end{align}
where \eqref{eq:defmi} follows from the definition of mutual information, 
\eqref{eq:indep_disagreegradset} follows since $\disagreegradset$ is independent from $\locgradset$ and \eqref{eq:uniform} holds since $\disagreegradset$ is a uniform selection of $\localcomp = \lfloor \nmalicious / \nhonest \rfloor$ indices out of $[\ngrad / \ngroup]$.
    To transmit the information in $\wtmdata$ with zero error to the \master, the workers need to send at least $\log_\alphasize \binom{\ngrad / \ngroup}{\lfloor \nmalicious / \nhonest \rfloor}$ symbols from $\galpha$.

  \FloatBarrier
  \subsection{Proof of \cref{thm:scheme}}
  \label{app:scheme_proof}
  In order to show the correctness of our scheme, we recall the following facts from before. The elimination tournament runs as long as there are contradicting responses among the non-eliminated workers. In each iteration of the elimination tournament, the main node eliminates at least one malicious worker (either by majority vote or by local computation). As soon as less than $\nhonest$ malicious workers are left, the elimination terminates. Therefore, the elimination tournament is guaranteed to terminate after at most $\nmalicious+1-\nhonest$ iterations. W.l.o.g. we consider $1 \leq \nhonest \leq \nmalicious+1$ here. For $\nhonest \geq \nmalicious+1$, the elimination tournament will terminate immediately for the reason explained before. Furthermore, honest workers are never eliminated, since they always respond with a correct value and never commit to an incorrect value. Having at least $\nhonest \leq 1$ honest workers in each fractional repetition group, the \master can recover the correct group result after the elimination tournament is terminated, and hence, for the output it always holds that $\gestim=\tgrad$. After having shown that our scheme is an \BGC scheme, we show the tuple $\big( \nround,\localcomp,\replfact,\commoh \big)$ in the remainder.
    
We start with the number of locally computed gradients $\localcomp$.
As explained above, each local computation of a partial gradient eliminates at least $\nhonest$ malicious workers in our scheme. The procedure halts when there are no more discrepancies among the initial responses of the non-eliminated workers, which is at latest when all $\nmalicious$ malicious workers are identified.
Therefore, the number of gradients computed locally is
\begin{equation*}
    \localcomp \leq \compbound.
\end{equation*}

Next, we analyze the communication rate $\commoh$. Every match causes each of the two competing workers to send up to $\left\lceil \log_2\left(\frac{\ngrad}{\ngroup}\right) \right\rceil$ symbols. The reason is that a competing worker sends one symbol for each node on a specific path of the match tree. The number of nodes in such a path is upper bounded by the height of the tree, which is $\left\lceil \log_2\left(\frac{\ngrad}{\ngroup}\right) \right\rceil$.
We have at most $\nmalicious+1-\nhonest$ matches. %
That is, the total amount of symbols transmitted for all matches is at most $2 \left(\nmalicious+1-\nhonest\right) \left\lceil \log_2\left(\frac{\ngrad}{\ngroup}\right) \right\rceil$.
Finally, each committing round after a match causes an additional communication load. Every worker of the competing consistent worker subsets transmits one bit, indicating whether or not the worker commits to the response of its representative. The communication load caused by committing is maximized by; first, maximizing the number of matches; and second, maximizing the number of involved nodes per match. This is achieved at the same time in the case where there are only two large subsets of consistent workers: the first consisting of the $\nhonest$ honest workers, and the second consisting of the $\nmalicious$ malicious workers. Furthermore, the malicious workers never commit to a representative's malicious value. Hence, we have $\nmalicious+1-\nhonest$ matches. The first match $\nmalicious+\nhonest$ has one-bit voting messages. Since at least one malicious worker can be eliminated, the number of voting messages will decrease by at least one per match accordingly. Therefore, the number of voting messages in the $l$-th match is given by $\nhonest+\nmalicious-(l-1)$. Summing over all $l=1,\dots,\nmalicious+1-\nhonest$, we can bound the total number of one-bit voting messages by $\frac{\left( \nmalicious+1-\nhonest \right)\left(\nmalicious+3\nhonest\right)}{2}$.
In terms of symbols from $\galpha$, this is a load of $\frac{\left( \nmalicious+1-\nhonest \right)\left(\nmalicious+3\nhonest\right)}{2\log_2{\card{\galpha}}}$.
In total, we obtain
\begin{equation*}
    \commoh \leq \left(\nmalicious+1-\nhonest\right) \left( 2 \left\lceil \log_2\left( \frac{\ngrad}{\ngroup}\right) \right\rceil+ \frac{\nmalicious+3\nhonest}{2\log_2{\card{\galpha}}} \right).
\end{equation*}
 
Finally, we analyze the number of communication rounds $\nround$. The number of communication rounds in a match is again upper bounded by the height of the tree, i.e., $\left\lceil \log_2\left(\frac{\ngrad}{\ngroup}\right) \right\rceil$.
Matches in different fractional repetition groups can be performed concurrently in the same rounds of the interactive protocol.
As explained before, there can be up to $\nmalicious$ matches. Again, if all malicious workers are in the same group, and if there are only two consistent worker subsets in this group of size $\nhonest$ and $\nmalicious$, respectively, then all $\nmalicious+1-\nhonest$ matches have to be executed sequentially. Resolving all conflicts, thus, requires
\begin{equation*}
    \nround \leq \left(\nmalicious+1-\nhonest\right) \left\lceil \log_2\left(\frac{\ngrad}{\ngroup}\right) \right\rceil.
\end{equation*}

  \FloatBarrier
  \subsection{Achievability vs. Converse for $\commoh$}
  \label{app:achievability_vs_converse}
  For convenience, we reproduce our achievability and converse bound on $\commoh$ from \cref{thm:scheme} and \cref{thm:converse_T}, respectively.
For fractional repetition data allocation, with $\ngroup$ groups, $\nworker = \ngroup(\nmalicious+\nhonest)$ workers, $\nmalicious$ of which are malicious, our scheme requires data transmission of at most
\begin{align*}
  \commoh[achieve] = \left(\nmalicious+1-\nhonest\right) \left( 2 \left\lceil \log_2\left( \frac{\ngrad}{\ngroup}\right) \right\rceil+ \frac{\nmalicious+3\nhonest}{2\log_2{\card{\galpha}}} \right)  
\end{align*}
during the interactive protocol as measured in symbols from $\galpha$.
Our lower bound shows, that for any computation optimal \BGC, \commoh is at least
\begin{align*}
  \commoh[bound] = {\log_{\alphasize} \binom{\ngrad/\ngroup}{\lfloor \nmalicious / \nhonest \rfloor}}.%
\end{align*}

\begin{figure}[h]
    \centering
    \resizebox{0.98\linewidth}{!}{
    \input{tikz/kappa_vs_p}
    }
    \caption{Comparison of converse and achievability for $\commoh$ over the dataset size $\ngrad$. We consider a system of $\nworker=10$ workers, $\ngroup=1$ group and an alphabet size $|\galpha|=2^{16}$. As the percentage of malicious workers rises from \SI{50}{\percent} to \SI{90}{\percent} the communication overhead of the scheme increases. }
    \label{fig:achievability_vs_converse_datacolumns}
\end{figure}

The gap between our scheme and the bound is depicted in~\cref{fig:achievability_vs_converse_datacolumns}.
It shows that for any parameter $\nmalicious$, the scheme is a constant factor away from the bound.
Typically, in distributed gradient descent applications the number of parameters $\graddim$ and the number of samples $\ngrad$ are very large, whereas the number of workers $\nworker$ and as a consequence $\nmalicious$, $\nhonest$ and $\ngroup$ are small by comparison.
For large numbers of samples $\ngrad$, the ratio $\commoh[achieve]/\commoh[bound]$ tends to 
\begin{align}
  \label{eq:conv_limit}
\lim_{\ngrad \to \infty} \frac{\commoh[achieve]}{\commoh[bound]} &= 2 \log_2(|\galpha|) \frac{(\nmalicious - \nhonest + 1)}{\lfloor \nmalicious / \nhonest 
\rfloor}.
\end{align}

The convergence behavior can be observed in \cref{fig:convergence}.
\begin{figure}[h]
    \centering
    \resizebox{0.98\linewidth}{!}{
    \input{tikz/kappa_relative_vs_p}
    }
    \caption{Convergence of the ratio $\frac{\commoh[achieve]}{\commoh[bound]}$ to the limit given in $\eqref{eq:conv_limit}$ for large numbers of samples.
      The parameters are $\nworker=10$, $\ngroup=10$, $|\galpha|=2^{16}$.
    For $\nmalicious=5$ and $\nmalicious=9$ the limits as in \eqref{eq:conv_limit} yield the same value.}
    \label{fig:convergence}
\end{figure}
Note that depending on the alphabet the communication overhead of our scheme can be slightly improved as stated in the following.
\begin{remark}[Compression Beyond the Alphabet Size]
  If for every pair of elements $a, b \in \galpha$ there exists a
  function $f: \galpha \mapsto \mathcal{B}$, that maps from $\galpha$ to a
smaller alphabet $\mathcal{B}$, such that $f(a) \neq f(b)$ and an operation $\oplus$ with the property $\forall c, d, e, g \in \galpha: f(c+d) \neq f(e+g) \implies f(c) \neq f(d)\text{ or }f(e) \neq f(g)$, then 
our scheme can be improved to $\commoh \geq \left(\nmalicious+1-\nhonest\right) \left( 2 \lceil \log_{|\galpha|}(|\mathcal{B}|)\rceil \left\lceil \log_2\left( \frac{\ngrad}{\ngroup}\right) \right\rceil+ \frac{\nmalicious+3\nhonest}{2\log_2{\card{\galpha}}} \right)$.
At the start of each match, the main node chooses not only the index $\compind$ but also the appropriate function $f$ and communicates it to the two workers. They then use $f$ to compress their transmitted symbols during the match. 
\end{remark}

  \pagebreak
  \subsection{Notation}
  \label{app:notations}
  \begin{table}[h]
\begin{tabularx}{\linewidth}{cX}
$\nworker$ & the number of workers \\
$\nmalicious$ & the number of malicious workers \\
$\ngroup$ & the number of groups in case of fractional repetition data allocation \\
$\nhonest$ & the number of guaranteed honest workers per group in case of fractional repetition data allocation \\
$\ngrad$ & the number of partial gradients \\
$\galpha$ & the message alphabet \\
$\tgrad$ & true gradient of the loss function over the full data set (total gradient) \\
$\gestim$ & estimated gradient of the loss function over the full data set (total gradient) \\
$\pgrad$ & true gradient of the loss function for sample $\sample$ (partial gradient) \\
$\wpgrad{\gind}{\wind}$ & claimed value of gradient $\pgrad$ from worker $\wind$ according to the attack strategy detailed in \cref{sec:attack} \\
$\allocmat$ & data allocation matrix of size $\ngrad \times \nworker$ \\
$\encfunset$ & list of encoding functions \\
$\encfun[\wind,\encind]$ & $\encind$-th encoding function available to worker $\wind$ \\
$\decfun$ & decoding function used by the \master \\
$\proto$ & interactive protocol \\
$\replfact$ & replication factor per partial gradient \\
$\commoh$ & communication overhead during the interactive protocol $\proto$ \\
$\localcomp$ & the number of local computations at the \master \\
$\nround$ & the number of rounds of the interactive protocol $\proto$ \\
$\roundind$ & the round index of the interactive protocol $\proto$ \\
$\gditer$ & the round index of the gradient descent iteration \\
$\recresset_{\roundind}$ & list of potentially corrupted worker responses in round $\roundind$ \\
$\irespset_{\roundind}$ & list of uncorrupted worker responses in round $\roundind$ \\
$\lgindset_{\roundind}$ & list of indices of partial gradients that are locally computed at the \master in round $\roundind$ \\
$\locgradset_{\roundind}$ & list of values of partial gradients that are locally computed at the \master in round $\roundind$ \\
$\disagreegradset$ & set of gradient indices on which workers disagree according to the attack strategy detailed in \cref{sec:attack}
\end{tabularx}
\vspace{1em}
\caption{Main notation.}
\end{table}

  \subsection{Algorithms}\label{app:algorithms}
  \balance
  \begin{algorithm}[h]
    \SetAlgoLined
    \SetKwInOut{Input}{Input}
    \SetKwInOut{Output}{Output}
    \SetKwInOut{Require}{Require}
    \Input{Workers $\worker[\indone]$ and $\worker[\indtwo]$, s.t. 
        $\noisyiresp_{0,\indone} \neq \noisyiresp_{0,\indtwo}$.
    }
    \Output{Values $\wpgrad{\indcheck}{\indone},\wpgrad{\indcheck}{\indtwo}$ and index $\indcheck$.}
    $\gind_\mathrm{min} \gets 1$\;
    $\gind_\mathrm{max} \gets \ngrad$\;
    $\compind \in \left\{ \compind^\prime \,\middle\vert\, \comp[\compind^\prime]{\noisyiresp_{0,\indone}} \neq \comp[\compind^\prime]{\noisyiresp_{0,\indtwo}} \right\}$\;
    \While{$\gind_\mathrm{max} - \gind_\mathrm{min} > 0$}{
        $\gind_\mathrm{half} \gets \gind_\mathrm{min} + \lceil \frac{\gind_\mathrm{max} - \gind_\mathrm{min}}{2} \rceil$\;
        $\textbf{request } \iresp_{\roundind,\indone} \gets
            \sum_{\gind=\gind_\mathrm{min}}^{\gind_{\mathrm{half}}}
            \comp{ \wpgrad{\gind}{\indone} }$ from $\worker[\indone]$\;
        $\textbf{request } \iresp_{\roundind,\indtwo} \gets 
            \sum_{\gind=\gind_\mathrm{min}}^{\gind_{\mathrm{half}}}
            \comp{ \wpgrad{\gind}{\indtwo} }$ from $\worker[\indtwo]$\;
        
        \eIf{$\noisyiresp_{\roundind,\indone} = \noisyiresp_{\roundind,\indtwo}$}{
            $\gind_\mathrm{min} \gets \gind_\mathrm{half}$\;
        }{
            $\gind_\mathrm{max} \gets \gind_\mathrm{half}$\;
        }
    
        $\roundind \gets \roundind + 1$\;
    }
    $\indcheck \gets \gind_\mathrm{min}$\;
    
    \caption{Match between two workers.}
    \label{alg:match}
\end{algorithm}

\begin{algorithm}[b]
    \newcommand{\groupset}{\ensuremath{\mathcal{G}}}
    \newcommand{\voteset}{\ensuremath{\mathcal{V}}}
    \SetAlgoLined
    \SetKwInOut{Input}{Input}
    \SetKwInOut{Output}{Output}
    \SetKwInOut{Require}{Require}
    \Input{Set $\groupset$ of disjoint worker groups, where $\workerset \in \groupset$ has $\workerset \subset \range{\nworker}$ and $\nhonest \leq \card{\workerset} \leq \nmalicious$.}
    \Require{$draw()$, $commit()$, $localComp()$.}
    \Output{Set $\elimset$ of malicious workers.}
    
    $\elimset \gets \left\{ 1,2,\dots,\nworker \right\} \setminus \cup_{\workerset \in \groupset} \workerset$\;
    \While{$\card{\groupset} > 1$}{
        $\workerset_1 \gets draw\left(\groupset\right)$;
        $\worker[1] \gets draw\left(\workerset_1\right)$\;
        $\workerset_2 \gets draw\left(\groupset \setminus \left\{ \workerset_1 \right\}\right)$;
        $\worker[2] \gets draw\left(\workerset_2\right)$\;
        $\comp{\wpgrad{\gind}{1}},\comp{\wpgrad{\gind}{2}},\gind,\compind \gets match(\worker[1],\worker[2])$ (\texttt{\cref{alg:match}})\;
        $\voteset_1 \gets commit\big( \comp{\wpgrad{\gind}{1}},\gind,\compind,\workerset_1 \big)$\;
        $\voteset_2 \gets commit\big( \comp{\wpgrad{\gind}{2}},\gind,\compind,\workerset_2 \big)$\;
        \If{$\card{\voteset_1} < \nhonest$}{
            $\elimset \gets \elimset \cup \voteset_1$\;
            $\workerset_1 \gets \workerset_1 \setminus \voteset_1$\;
        }
        \If{$\card{\voteset_2} < \nhonest$}{
            $\elimset \gets \elimset \cup \voteset_2$\;
            $\workerset_2 \gets \workerset_2 \setminus \voteset_2$\;
        }
        \If{$\card{\voteset_1} \geq \nhonest$ and $\card{\voteset_2} \geq \nhonest$}{
            $\comp{\pgrad[\gind]} \gets localComp(\gind,\compind)$\;
            \If{$\comp{\wpgrad{\gind}{1}} \neq \comp{\pgrad[\gind]}$}{
                $\elimset \gets \elimset \cup \voteset_1$\;
                $\workerset_1 \gets \workerset_1 \setminus \voteset_1$\;
            }
            \If{$\comp{\wpgrad{\gind}{2}} \neq \comp{\pgrad[\gind]}$}{
                $\elimset \gets \elimset \cup \voteset_2$\;
                $\workerset_2 \gets \workerset_2 \setminus \voteset_2$\;
            }
        }
        \For{$i=1,2$}{
            \If{$\card{\workerset_i} < \nhonest$}{
                $\groupset \gets \groupset \setminus \workerset_i$\;
            }
        }
    }
    \caption{Elimination Tournament.}
    \label{alg:elimination_tournament}
\end{algorithm}

}

\end{document}